\documentclass[a4,11pt]{article}

\usepackage{graphicx}

\usepackage{amsthm}
\usepackage{float}
\restylefloat{table}
\usepackage{amsmath}
\usepackage{amssymb}
\usepackage{amsfonts,cancel}
\usepackage[usenames]{color}
\usepackage{authblk} 
\usepackage{soul} 

\newcommand{\be}{\begin{equation}}
\newcommand{\ee}{\end{equation}}
\newcommand{\bea}{\begin{eqnarray}}
\newcommand{\eea}{\end{eqnarray}}
\newcommand{\Kdt}{{\hbox{\tiny K}}}
\newcommand{\rovno}{\!\!\!& = &\!\!\!} 
\newcommand{\eqdef}{\!\!\!& \equiv &\!\!\!}
\newcommand{\X}{{\rho}}

\def \d {{\rm d}}
\def \dd {{\rm d}}

\def \pul {\textstyle{\frac{1}{2}}}

\def \H {\mathcal{H}}
\def \T {\mathcal{T}}
\def \B {\mathcal{B}}

\newcommand{\beqn}{\begin{eqnarray}}
\newcommand{\eeqn}{\end{eqnarray}}
\newcommand{\pa}{\partial}

\usepackage{amsmath,amsthm,amssymb,cite,enumerate}

\def \d {{\rm d}}

\usepackage{amsmath}

\begin{document}

\title{Topological black holes in  higher derivative gravity}

\author{Alena Pravdov\' a\thanks{pravdova.math.cas.cz}}

\author{Vojt\v ech Pravda\thanks{pravda.math.cas.cz}}

\author{Marcello Ortaggio\thanks{ortaggio(at)math(dot)cas(dot)cz}}

\affil{Institute of Mathematics of the Czech Academy of Sciences, \newline \v Zitn\' a 25, 115 67 Prague 1, Czech Republic}

\maketitle

\begin{abstract}

We study static black holes in quadratic gravity with planar and hyperbolic symmetry and non-extremal horizons. We obtain a solution in terms of an infinite power-series expansion around the horizon, which is characterized by two independent integration constants -- the black hole radius and the strength of the Bach tensor at the horizon. While in Einstein's gravity, such black holes require a negative cosmological constant $\Lambda$, in quadratic gravity they can exist for any sign of  $\Lambda$ and also for $\Lambda=0$.  Different branches of Schwarzschild-Bach-(A)dS or purely Bachian black holes are identified which admit distinct Einstein limits. Depending on the curvature of the transverse space and the value of $\Lambda$, these Einstein limits result in  (A)dS-Schwarzschild spacetimes with a transverse space of arbitrary curvature (such as black holes and naked singularities) or in Kundt metrics of the (anti-)Nariai type (i.e., dS$_2\times$S$^2$, AdS$_2\times$H$^2$, and flat spacetime).  In the special case of toroidal black holes with $\Lambda=0$, we also discuss how the Bach parameter needs to be fine-tuned to ensure that the metric does not blow up near infinity and instead matches asymptotically a Ricci-flat solution.

\end{abstract}

\section{Introduction}

Black holes can be regarded as the most fundamental objects in
gravity, serving as theoretical laboratories to study various aspects of gravitational theories. 
In general relativity, Hawking's theorem states that spacelike cross sections of an event horizon in a stationary asymptotically flat spacetime are topologically 2-spheres, assuming also that the dominant energy condition holds \cite{Hawking72,HawEll73}. By relaxing some of the assumptions in Hawking's theorem, one can obtain more general horizon geometries. For instance, in a locally asymptotically anti-de Sitter (AdS) spacetime (where the asymptotic flatness and dominant energy condition are both violated), it is possible to construct topological black holes for which the spacelike cross section of the event horizon can be a compact Riemann surface of any genus~$g$ \cite{Lemos95,HuaLia95,AmiBenHolPel96,Mann97,Vanzo97,BriLouPel97}.

Stemming from the early results of \cite{Stelle77,Stelle78}, recently there has been a great interest in the study of static, spherically symmetric black holes in quadratic gravity \cite{Luetal15,Luetal15b,Luetal17,KokKonZhi17,Goldstein18,Podolskyetal18,Svarcetal18,BonSil19,Podolskyetal20,Pravdaetal21,Huang22}, where corrections quadratic in the curvature are  added to the Einstein-Hilbert action 
\be
S {=\int \d^4 x\, \sqrt{-g}{\cal L}}= \int \d^4 x\, \sqrt{-g}\, \Big(
\gamma \,(R-2\Lambda) +\beta\,R^2  - \alpha\, C_{abcd}\, C^{abcd}
\Big)\,,
\label{action}
\ee
where ${\gamma=1/(16\pi G)}$, $G$ is the Newtonian constant (we will use units such that $G=1=c$), $\Lambda $ is the cosmological constant, and  $\alpha$, $\beta$  are {coupling} constants of the theory.

It is well known \cite{Buchdahl48_2,Buchdahl48_3} that all Einstein spaces $R_{ab}=\Lambda g_{ab}$  automatically solve the vacuum field equations of quadratic gravity. Vacuum black holes appearing in Einstein's gravity are thus in a sense trivial solutions to quadratic gravity. 
However, recently  it has been shown  \cite{Luetal15,Luetal15b}  that, besides the standard  Schwarzschild black hole,  quadratic gravity also admits another static, spherically symmetric black hole solution over and above Schwarzschild. Extensions of such black holes with a non-vanishing cosmological constant $\Lambda$ have been studied in \cite{Svarcetal18,Pravdaetal21}.
Although these non-Schwarzschild (or ``Schwarzschild-Bach'') black holes of \cite{Luetal15,Luetal15b} and \cite{Svarcetal18,Pravdaetal21} have nontrivial Ricci tensor, the  Ricci scalar $R$  is vanishing or constant, respectively. In fact, the  Ricci scalar  is constrained
by the trace no-hair theorem of \cite{Nelson2010,Luetal15b} which states that for static, spherically symmetric black holes in quadratic gravity, the Ricci scalar  is either zero (in the asymptotically flat case)
or constant (assuming that  $R$ is sufficiently quickly approaching a
constant at infinity) throughout the spacetime. Furthermore, for $R=$const, the field equations of quadratic gravity considerably simplify to (assuming in~\eqref{action} $\gamma\neq0$)
\be
R_{ab}-\Lambda \, g_{ab}=4k\, B_{ab}\,,
\label{eq:feq}
\ee
where $B_{ab}$ is the Bach tensor \eqref{defBach} and the constant $k$ is defined by \eqref{k}.

The Schwarzschild-Bach black holes  are thus clearly  distinguished by a non-vanishing Bach tensor. In fact, it has been shown that for these black holes, the vanishing of the Bach tensor on the horizon guarantees the vanishing of the Bach tensor throughout the spacetime \cite{Podolskyetal18,Podolskyetal20,Svarcetal18,Pravdaetal21}. 
These black holes are therefore characterized by two parameters, namely, the radius of the black hole $\bar r_0$ and a Bach parameter $b$ (denoted as $\delta$ in some of the literature) measuring a deviation from the Schwarzschild solution and related to the value of the Bach invariant $B_{ab} B^{ab}$ on the horizon.
An exact solution in a closed form describing this black hole is unknown -- in the $\Lambda=0$ case, evidence for the existence of this black hole has been provided  in \cite{Luetal15,Luetal15b} by taking the first few terms in the near-horizon expansion and numerically integrating the solution out from the horizon to some point outside the horizon, before the numerical solution diverges (the fourth order equations of motion are
numerically unstable).  As it turns out, to ensure the asymptotic flatness of the spacetime (such as to kill the growing Yukawa modes), one needs to fine-tune the parameters $\bar r_0$ and $b$ using numerical methods, thus effectively ending up with a one-parameter family of solutions \cite{Luetal15,Luetal15b,Luetal17,Goldstein18,BonSil19,Huang22}.  Very recently, it has been shown that for a given $\bar r_0$  there exist at least two values of $b$ giving an asymptotically flat black hole \cite{Huang22}. However, introducing a non-zero cosmological constant or non-spherical horizon topologies into the picture may have significant consequences on the physics of these black holes. Various results for the case $\Lambda\neq0$ have been obtained in \cite{Luetal12,FanLu15,Svarcetal18,Pravdaetal21}.\footnote{More results, including several exact solutions, are available in special quadratic theories such as conformal or pure $R^2$ gravity for spherical \cite{Buchdahl53,Riegert84,ManKaz89,DesTek03cqg} and topological \cite{Klemm98_cqg,CaiLiuSun09,Cognolaetal11,Luetal12,Cognolaetal15} black holes. These special theories will not be considered in the present paper.}

In this work, we will broaden the search for black holes in quadratic gravity. In addition to static black holes with spherical symmetry, we will also include hyperbolic and planar symmetry.
The corresponding metric ansatz thus reads
\be
\dd s^2 = -h(\bar r)\,\dd t^2+\frac{\dd \bar r^2}{f(\bar r)}+\bar r^2 \dd \omega^2, \ \ \  \ 
\dd\omega^2=\frac{\dd x^2+\dd y^2}{\left(1+\frac{\epsilon}{4} (x^2+y^2)\right)^2} , \label{physmet}
\ee
where the transverse geometry is $S^2$, $E^2$, and $H^2$ for $\epsilon=$ $+1$, $0$, and $-1$, respectively. As in Einstein's gravity, one can use the metric \eqref{physmet} with $\epsilon=$ $0$ or $-1$ to construct topological black holes, for which the horizon is a flat torus ($g=1$) or a Riemann surface of genus $g>1$, respectively (such compactifications are discussed in \cite{Lemos95,HuaLia95,AmiBenHolPel96,Mann97,Vanzo97,BriLouPel97}).

It is well known that in Einstein's gravity,  black holes with $\epsilon\le0$ require $\Lambda<0$. Here we will show  that these constraints do not apply to quadratic gravity and that such black holes can exist for  any sign of  $\Lambda$ as well  as for $\Lambda=0$.

We will take the  conformal-to-Kundt approach, recently employed in  \cite{Podolskyetal18,Podolskyetal20} and \cite{Svarcetal18,Pravdaetal21}, to study static, spherically symmetric black holes in quadratic gravity with vanishing and nonvanishing $\Lambda$, respectively. Accordingly, the standard metric \eqref{physmet} is rewritten in a form conformal to the Kundt metric. This greatly simplifies the field equations of quadratic gravity,  at the price of working in somewhat physically less transparent coordinates.
Together with the assumption $R=$const, motivated by the trace no-hair theorem, the resulting simplification 
of the field equations will enable us to obtain recurrent formulas for  series coefficients of the metric functions in power-series expansions, and thus to include also analytical results in the study of these black holes. 

In section~\ref{sec_backg}, we present necessary background material, such as the field equations of quadratic gravity following from the action \eqref{action}, and the conformal-to-Kundt approach to simplify them.

In section~\ref{sec_FEQ}, the field equations are derived for the ansatz~\eqref{physmet} reexpressed in the Kundt coordinates. We then use a Frobenius-like approach to solve the equations in the vicinity of a generic hypersurface of a constant radius. We use infinite power-series expansions and the  indicial equations to determine the possible leading powers of solutions. In particular, some of these solutions admit extremal or non-extremal horizons.

Section~\ref{sec_BH_solutions} then focuses on the study of  solutions with non-extremal horizons, which is the main focus of the present paper. The recurrent formulas for the series coefficients are determined. Interestingly, depending on $\Lambda$ and $\epsilon$, the Einstein limit of the black hole solutions constructed here contains not only (A)dS-Schwarzschild black holes with a transverse space of arbitrary curvature, but also naked singularities,  Kundt metrics of the {(anti-)}Nariai type -- dS$_2\times$S$^2$, AdS$_2\times$H$^2$, and  flat spacetime. As in the spherical case \cite{Luetal15,Luetal15b,Luetal17,Goldstein18,Podolskyetal18,Svarcetal18,BonSil19,Podolskyetal20,Pravdaetal21,Huang22}, one might expect that, by fine-tuning the Bach parameter, a quadratic-gravity black hole with $\Lambda=0$ will asymptote an (appropriate) Ricci-flat metric near spatial infinity. We will give evidence to support this expectation  by  fine-tuning  a planar black-hole solution ($\epsilon=0$) with vanishing $\Lambda$, using the polynomial expansion.\footnote{However, more solid evidence would be to match the expansion in the vicinity of the horizon with the asymptotic expansion in the form of logarithmic-exponential transseries  (cf.~\cite{Goldstein18}).} In contrast, fine-tuning is not necessary to obtain an asymptotically Einstein spacetime in the case of nonvanishing $\Lambda$ within a  certain continuous range of parameters of the solution (see also sec. \ref{sec_concl}).

Concluding comments are given in section~\ref{sec_concl}, also on some related results obtained for the $\Lambda<0$ case in \cite{Luetal12,FanLu15}.

\section{Background}
\label{sec_backg}

\subsection{Quadratic gravity}

\label{subsec_QG}

The field equations following from action \eqref{action} read
\be
\gamma \left(R_{ab} - {\pul} R\, g_{ab}+\Lambda\,g_{ab}\right)-4 \alpha\,B_{ab}
+2\beta\left(R_{ab}-\tfrac{1}{4}R\, g_{ab}+ g_{ab}\, \Box - \nabla_b \nabla_a\right) R = 0 \,, \label{fieldeqsEW}
\ee
where $B_{ab}$ is the {Bach tensor}
\be
B_{ab} \equiv \big( \nabla^c \nabla^d + {\pul} R^{cd} \big) C_{acbd} \ , \label{defBach}
\ee
which is traceless, symmetric, conserved, and well-behaved under a conformal transformation $g_{ab}=\Omega^2 \tilde g_{ab}$:
\begin{equation}
	g^{ab}B_{ab}=0 \,, \qquad B_{ab}=B_{ba} \,, \qquad
	\nabla^b B_{ab}=0
	\,, \qquad B_{ab}=\Omega^{-2}\tilde B_{ab}\,.
	\label{Bachproperties}
\end{equation}
{For four-dimensional (conformally) Einstein spacetimes, the Bach tensor vanishes identically \cite{Buchdahl53}.}

As discussed above, in this work we will restrict ourselves to solutions with $R=$const.\footnote{This is not a restriction for Einstein-Weyl gravity, i.e., when $\beta=0$, since it then clearly follows from~\eqref{fieldeqsEW}.} Then the trace of \eqref{fieldeqsEW}  reduces to $\gamma(R-4\Lambda)=0$. Assuming $\gamma\neq0$ (to keep the Einstein-Hilbert term in the action \eqref{action}) we thus have
	\be
	R=4\Lambda ,
	\label{R=const}
	\ee
	and the field equations~\eqref{fieldeqsEW} simplify to \eqref{eq:feq},
where we have defined 
	\be
	k\equiv\frac{\alpha}{\gamma+8\beta\Lambda} \qquad (\gamma+8\beta\Lambda\neq0) .
	\label{k}
	\ee

For later purposes, let us note that solutions with vanishing Bach tensor reduce to Einstein spacetimes (cf.~\eqref{eq:feq}). The latter are of constant curvature if, in addition, ${\H}'' +2\epsilon=0$ (cf.~\eqref{invC}). We excluded the specially fine-tuned case $\gamma+8\beta\Lambda=0$, for which the field equations (with~\eqref{R=const}) reduce to $\alpha B_{ab}=0$ (see, e.g., \cite{Pravdaetal17}), as in conformal gravity -- all static (spherical, hyperbolic, or planar) black holes are already known in this case \cite{Buchdahl53,Riegert84,ManKaz89,Klemm98_cqg}. If, in addition, also $\alpha=0$, one has a special Einstein-$R^2$ gravity for which any metric with $R=4\Lambda$ is a solution (see, e.g., \cite{DesTek03cqg,CaiLiuSun09,Ayon-Beatoetal10,HenEslMou12} for some examples).

\subsection{Conformal-to-Kundt ansatz}

\label{sec_BH metric}

We are interested in static black-hole solutions with spherical, hyperbolic, or planar symmetry. Instead of the standard Schwarzschild coordinates, throughout the paper, we will mostly employ the conformal-to-Kundt form of the metric introduced in~\cite{Pravdaetal17,Podolskyetal18,Podolskyetal20}. This enables one to describe such spacetimes in the form		 
\be
\dd s^2 \equiv \Omega^2(r) \,\dd s^2_\Kdt = \Omega^2(r)
\Big[\,\dd \omega^2 -2\,\dd u\,\dd r+\H(r)\,\dd u^2 \,\Big]\,,
\label{BHmetric}
\ee
for which the resulting field equations are considerably simpler.

The metric \eqref{BHmetric} admits a gauge freedom
\be
r \to \lambda\,r+\upsilon\,, \qquad u \to \lambda^{-1}\,u \,, 
\label{scalingfreedom}
\ee
where $\lambda\,, \upsilon$ are constants, i.e., it is invariant up to rescaling $\H\to\lambda^2\H$.

When $\Omega'\neq0\neq \H$ one can also define the standard Schwarzschild coordinates by \cite{Pravdaetal17}
\begin{equation}
\bar{r} = \Omega(r)\,, \qquad t = u - \int\! \frac{\dd r}{\H(r)} \,,
\label{to static}
\end{equation}
giving rise to the line-element \eqref{physmet} with
\begin{equation}
h = -\Omega^2\, \H , \quad f = -\left(\frac{\Omega'}{\Omega}\right)^2 \H  , 
\label{Schwarz}  
\end{equation}
where a prime denotes differentiation with respect to $r$.

\section{Field equations and classes of power-series solutions}
\label{sec_FEQ}

\subsection{Ricci, Weyl, and Bach tensors for the Kundt seed metric}

The  nontrivial Ricci tensor components and  the Ricci scalar of the Kundt background metric $\dd s^2_\Kdt$ of \eqref{BHmetric} read (cf. \cite{Podolskyetal20} for $\epsilon=1$ case)
\begin{eqnarray}
	R_{ru}^\Kdt \rovno -\pul\,{\cal H}'' \,, \qquad  \quad\ 	R_{uu}^\Kdt = -{\cal H}\,R_{ru}^\Kdt \,,\label{Ricci uu5} \\
	R_{xx}^\Kdt \rovno R_{yy}^\Kdt = \epsilon g^\Kdt_{xx} \,, \qquad 	R^\Kdt={\cal H}''  + 2 \epsilon \, .
\end{eqnarray}

The non-vanishing components of the Weyl and Bach tensors are, respectively,
\bea
	C_{ruru}^\Kdt \rovno {\textstyle -\frac{1}{6} R^\Kdt} \,, \qquad\qquad\qquad\ \ \  	C_{riuj}^\Kdt= {\textstyle \frac{1}{12} R^\Kdt \,g^\Kdt_{ij}} \,, \\
	C_{kilj}^\Kdt \rovno {\textstyle \frac{1}{6}R^\Kdt\,(g^\Kdt_{kl}g^\Kdt_{ij}-g^\Kdt_{kj}g^\Kdt_{il}) } \,, \quad C_{uiuj}^\Kdt =  -{\cal H}\, C^\Kdt_{riuj}  \, \label{WeylfK}
\eea
and 
\begin{eqnarray}
	B_{rr}^\Kdt \rovno {\textstyle -\frac{1}{6}\,{\cal H}'''' } \,, \label{Bach rr} \qquad 
	B_{ru}^\Kdt = {\textstyle \frac{1}{12}\,
		\big(2\,{\cal H}{\cal H}''''+{\cal H}'{\cal H}'''-{\textstyle\frac{1}{2}}{{\cal H}''}^2 +2\epsilon^2\big) } \,, \label{Bach ru}\\
	B_{uu}^\Kdt \rovno  -{\cal H}\,B_{ru}^\Kdt \,, \qquad 
	B_{xx}^\Kdt = B_{yy}^\Kdt = {\textstyle \frac{1}{12}}\,g^\Kdt_{xx}\,
	\big({\cal H}{\cal H}''''+{\cal H}'{\cal H}'''-{\textstyle\frac{1}{2}}{{\cal H}''}^2 +2\epsilon^2\big)   \,. \label{Bach xx}
\end{eqnarray}

\subsection{{Ricci and Bach tensors} for the full metric}

\label{subsec_curv}

The nontrivial Ricci tensor components and the Ricci scalar for the full metric~\eqref{BHmetric} are
\begin{eqnarray}
R_{rr} \rovno -2\Omega^{-2}\big(\Omega\Omega''-2{\Omega'}^2\big) \,,  \qquad 
R_{ru} = -\pul \Omega^{-2}\big(\Omega^2 {\cal H}\big)'' \,, \label{RT_R ru}\\
R_{uu} \rovno  -{\cal H}\, {R}_{ru} \,,   \qquad \qquad\qquad\quad\ \
R_{xx} =  {R}_{yy} = 
\Omega^{-2}g_{xx}^\Kdt \,
\big[ \big({\cal H}\Omega\Omega'\big)'+\epsilon\Omega^2 \big]   \,,\label{RT_R xx}\\
R \rovno 6\Omega^{-3} \big[({\cal H}\Omega')'
+{\textstyle \frac{1}{6}} ({\cal H}''+2 \epsilon )\Omega \big] \,.
\label{barR}
\end{eqnarray}

The Bach tensor of the full metric can be obtained by a rescaling \eqref{Bachproperties}
\be
B_{ab} = \Omega^{-2}B_{ab}^\Kdt\,, \label{OmBach}
\ee
while $C^a_{\phantom{a}bcd}=C^{\Kdt a}_{\phantom{Ka}bcd}$, as well known.

The Ricci squared, Bach, and Weyl invariants read\footnote{Note that the field equations~\eqref{eq:feq} have been used to obtain \eqref{invR}.}
\begin{align}
	R_{ab}\, R^{ab} &=  4\Lambda^2+16k^2\, B_{ab} B^{ab} \,, \label{invR}\\
	B_{ab}\, B^{ab} &=  \tfrac{1}{72}\,\Omega^{-8}\,\big[(\B_1)^2 + 2(\B_1+\B_2)^2\big] \,,\label{invB}\\
	C_{abcd}\, C^{abcd} &=  \tfrac{1}{3}\,\Omega^{-4}\,\big({\H}'' +2\epsilon\big)^2 \,, \label{invC}
\end{align}
where
the two independent components of the Bach tensor, $\B_1(r)$ and $\B_2(r)$, are 
\be
	\B_1 \equiv {\H}{\H}''''\,, \qquad \B_2 \equiv {\H}'{\H}'''-\tfrac{1}{2}{{\H}''}^2 +2\epsilon^2\,. \label{B2}
\ee

It is useful to note that $B_{ab}=0\Leftrightarrow\B_1=0=\B_2$. It can be also verified easily that $C_{abcd}=0\Leftrightarrow {\H}'' +2\epsilon=0$.

\subsection{Derivation and simplification of the field equations}

\label{FEQ}

Following \cite{Podolskyetal20,Pravdaetal21}, in this section, we show that the field equations \eqref{eq:feq} for the full metric \eqref{BHmetric} reduce to two coupled autonomous nonlinear differential equations.

The  nontrivial components $rr$, $ru$, and $xx$ of  the field equations \eqref{eq:feq} read 
\begin{align}
	\Omega\Omega''-2{\Omega'}^2 & = \tfrac{1}{3}k\, {\H}'''' \,, \label{Neq_rr} \\
	\big(\Omega^2 {\H}\big)'' -2\Lambda \Omega^4 & = -\tfrac{2}{3}k \big(2\,{\H}{\H}''''+{\H}'{\H}'''
	-{\textstyle\frac{1}{2}}{{\H}''}^2 +2\epsilon^2\big) \,, \label{Neq_ru} \\
	\big({\H}\Omega\Omega'\big)'+\epsilon\Omega^2 -\Lambda \Omega^4 & = \tfrac{1}{3}k \,\big({\H}{\H}''''+{\H}'{\H}'''
	-{\textstyle\frac{1}{2}}{{\H}''}^2 +2\epsilon^2 \big) \,. \label{Neq_xx}
\end{align}
The $yy$ component of \eqref{eq:feq} is identical to the $xx$ component and the $uu$ component is a multiple of  the $ru$ component.

The trace~{\eqref{R=const}} of the field equations takes the form
\begin{equation}
	\T\equiv{\H}\Omega''+{\H}'\Omega'+{\textstyle \frac{1}{6}} ({\H}''+2\epsilon)\Omega =
	{\textstyle \frac{2}{3}\Lambda \,\Omega^3 } \,.
	\label{traceC}
\end{equation}

As in \cite{Podolskyetal20,Pravdaetal21}, let us introduce a conserved ($\nabla^b J_{ab}\equiv 0$) symmetric tensor $J_{ab}$ 
\be
J_{ab}\equiv  R_{ab}-\pul  R\, g_{ab}+\Lambda \,  g_{ab} - 4k\,  B_{ab}\,.
\label{defJab}
\ee
The non-trivial components are
$	J_{rr}$,
$	J_{uu}=-{\H}\, J_{ru} $, and
$	J_{xx}= J_{yy}  $.
The vacuum field equations  \eqref{fieldeqsEW}, assuming $R=$const, then  take the form
$J_{ab}=0$. 
When $\Omega'\not=0$, one can show that once the field equations $J_{rr}=0$ and $J_{ru}=0$ hold, 
 then also $J_{xx}$ vanishes (see Appendix C in  \cite{Pravdaetal21} for a more detailed discussion in the $\epsilon=1$ case).\footnote{In this paper, we will not study non-Einstein spacetimes with $\Omega=$const, which correspond to Kundt metrics.}

The first field equation  $J_{rr}=0$ reduces to \eqref{Neq_rr}. Substituting for  ${{\H}''''}$ from \eqref{Neq_rr},
the equation $J_{ur}=0$ can be simplified and we arrive at the following final form of the field equations
\begin{align}
	\Omega\Omega''-2{\Omega'}^2 = &\ \tfrac{1}{3}k\,{\H}'''' \,, \label{Eq1C}\\
	\Omega\Omega'{\H}'+3\Omega'^2{\H}+\epsilon\Omega^2 -\Lambda \Omega^4
	= &\ \tfrac{1}{3}k \big({\H}'{\H}'''-{\textstyle\frac{1}{2}}{{\H}''}^2 +2\epsilon^2 \big)\,. \label{Eq2C}
\end{align}

\subsection{Classes of power-series solutions}

Let us assume that  the metric functions $\Omega(r) $ and  $\H(r)$ in~\eqref{BHmetric} can be expanded as infinite power series in ${\Delta \equiv r-r_0}$ around a hypersurface $r=r_0$, i.e.,
\be
	\Omega(r)=\Delta^n   \sum_{i=0}^\infty a_i \,\Delta^{i}\,, \qquad 	\H(r)=\Delta^p \,\sum_{i=0}^\infty c_i \,\Delta^{i} . 
	\label{rozvojcalH0}
\ee
Substituting these expansions into the field equations \eqref{Eq1C} and \eqref{Eq2C} and comparing the leading terms leads  to constraints on  possible values of $n$ and $p$. In Table~\ref{table_cases}, we summarize the classes allowing for a vanishing $\Lambda$ or an arbitrary $\Lambda$. Note that there exist further classes allowing only for certain discrete nonzero values of $\Lambda$ -- these are not included in the table and will be studied elsewhere. The case $\epsilon=+1$ has been already analyzed in \cite{Podolskyetal20,Pravdaetal21} (including the discrete values of $\Lambda$).

\begin{table}[H]
	\begin{center}
		\begin{tabular}{|c||c|c||c||}
			\hline
			Case  &
			Class $[n,p]$ &
			$\Lambda$ &
			$\epsilon $  \\
			\hline
			I& 	$[-1,2]$ & 0 & $ \not=0$  \\
			& 	$[-1,3]$ & 0 & 0 \\	
			& 	$[0,p>2]$ & 0 & 0\\		
			\hline
			II& 	$[0,0]$ & any & any  \\
			& 	$[0,1]$ & any & any\\
			& 	$[1,0]$ & any & any \\
			\hline
			III& 	$[-1,0]$ & any & any  \\
			& 	$[0,2]$ & any &  $ \not=0$  \\
			\hline	  	  	  
		\end{tabular} \\[2mm]
	\caption{\small{Values of $[n,p]$ compatible with indicial equations following from \eqref{Eq1C}--\eqref{rozvojcalH0}.
	Certain further cases allowing only for discrete nonzero values of $\Lambda$ are not included in the table. See \cite{Podolskyetal18,Svarcetal18,Podolskyetal20,Pravdaetal21} for details in the case $\epsilon=+1$.}}
		\label{table_cases}
	\end{center}
\end{table}

In the rest of the paper, we will study the $[0,1]$ case, which corresponds to spacetimes admitting a non-extremal Killing horizon. For certain ranges of  parameters, this can be interpreted as a black hole horizon. The remaining cases (as well as additional classes obtained using asymptotic expansions in negative powers of $r$) will be studied elsewhere.

\section{Case $[0,1]$: black holes with a non-extremal horizon}

\label{sec_BH_solutions}

\subsection{Preliminaries}

From now on, we focus on solutions for which $n=0$ and $p=1$ in~\eqref{rozvojcalH0}. This means that we are expanding the metric near a non-extremal Killing horizon, located at $r=r_0$. Let us thus relabel
\be
 r_h\equiv r_0 .
\ee
Because of the freedom~\eqref{scalingfreedom}, the particular value $r_0$ has no physical meaning. However, in the physical coordinates~\eqref{to static}, \eqref{physmet}, the horizon radius is given by
\be
 \bar r_h\equiv\Omega(r_h)=a_0>0 ,
 \label{a_0}
\ee 
which is a dimensionful scale set by $a_0$ (which is effectively an integration constant, see the following for more comments). Without loss of generality, we have fixed the sign of $a_0$ using the invariance of~\eqref{BHmetric} under $\Omega\to-\Omega$.

Before discussing the metric on-shell, let us note that, when $a_1\neq0$, the leading order behaviour of the metric functions $h$ and $f$ in the Schwarzschild coordinates~\eqref{physmet} is given by
\be
  h=-\frac{c_0a_{0}^2}{a_1}(\bar r-\bar r_h)+{\cal O}\left((\bar r-\bar r_h)^2\right) , \qquad f=-\frac{c_0a_1}{a_0^2}(\bar r-\bar r_h)+{\cal O}\left((\bar r-\bar r_h)^2\right) . \label{hf_Schw} 
\ee
In order to have an outer black hole horizon, we thus need to take 
	\be
		c_0 a_1<0\,,\label{hor_conds}
		\ee 
which ensures that both $h$ and $f$ are positive in the exterior region $\bar r>\bar r_h$ in the vicinity of $\bar r=\bar r_h$ (negative $h$ and $f$ would correspond, e.g., to an inner or a cosmological horizon). Note that, when  $a_1>0$,\footnote{This can be always achieved using	the gauge transformation~\eqref{scalingfreedom}. The special case $a_1=0$ leads to fractional steps in the expansion of the metric functions $f(\bar r)$ and $h(\bar r)$ in the physical coordinates~\eqref{physmet}, and will not be considered in the following (see \cite{Podolskyetal20,Pravdaetal21} for the case $\epsilon=+1$).} near and across the horizon $\bar r$ is monotonically increasing with $r$, while for $a_1<0$, $\bar r$ is monotonically decreasing with $r$. Therefore $\pa_r$ is outward/inward according to the sign of $a_1$.

\subsection{General solution}

\label{subsec_general}

The lowest nontrivial order of the trace equation \eqref{traceC} gives
\begin{align}
	a_1=\frac{a_0}{3c_0}\left[2\Lambda a_0^2-(\epsilon+c_1)\right] \,,
	\label{nonSchwinitcond3}
\end{align}
and then the lowest nontrivial order of  \eqref{Eq2C} implies
\be
c_2 =\frac{1}{6kc_0}\left[2k(c_1^2-\epsilon^2)+a_0^2(2\epsilon-c_1-\Lambda a_0^2)\right]\,.
\label{nonSchwinitcond2}
\ee

At any arbitrary higher order, one finds that the $[0,1]$ solution  to eqs. \eqref{Eq1C}, \eqref{Eq2C} is given by the recurrent formulas
 \bea
 c_{l+2}\!\!\!&=&\!\!\!\frac{3}{k\,(l+3)(l+2)(l+1)l}\,\sum^{l}_{i=0}a_i \,
 a_{l+1-i}(l+1-i)(l-3i)  \qquad \forall\ l\ge 1\,,
 \label{nonSchwinitcondc}\\
 a_{l}\!\!\!&=&\!\!\!\frac{1}{l^2c_0}\Bigg[\tfrac{2}{3}\Lambda \sum^{l-1}_{j=0}{a_{l-1-j}}\sum^{j}_{i=0}a_i\,a_{j-i}-\tfrac{1}{3}\,\epsilon a_{l-1}
 -\sum^{l}_{i=1}c_i\,a_{l-i}\left[l(l-i)+\tfrac{1}{6}i(i+1)\right]\Bigg] \ \ \
 \forall \ l\geq 2\,.  \label{nonSchwinitconda}
 \eea
The three parameters $a_0$, $c_0$, and $c_1$ remain arbitrary and can be thought of as integration constants.

It is also useful to observe that the Bach and Weyl invariants~(\ref{invB}), (\ref{invC}) at ${r=r_h}$ read
\bea
B_{ab}\,B^{ab}(r_h) &=& \left(  \frac{\epsilon^2-c_1^2+3c_0c_2}{3 a_0^4} \right)^2
=\left(\frac{c_1-2\epsilon+\Lambda a_0^2}{6k a_0^2}\right)^2
= \frac{b^2}{4 k^2 a_0^4 } \,,\ \label{BachInvariant}\\
C_{abcd}\, C^{abcd}(r_h)& =& \frac{4}{3 a_0^4}(\epsilon+ c_1)^2 \,,\label{BInv2}
\eea
where we have introduced a {dimensionless} Bach parameter $b$ by
\be
b \equiv \frac{1}{3}\left(c_1-2\epsilon+\Lambda a_0^2\right)\,, \label{b_definice}
\ee
which measures the strength of the Bach tensor at the horizon (it is proportional to $\B_2(r_h)$, see~\eqref{B2}).

For definiteness, using eqs.~\eqref{nonSchwinitcond3}, \eqref{nonSchwinitcond2} and the recurrent relations \eqref{nonSchwinitcondc}, \eqref{nonSchwinitconda},  the first few coefficients expressed in terms of free parameters $a_0$, $c_0$, and $b$ read
\begin{align}
	& a_{1} = -\frac{a_{0}}{c_{0}}\left[(\epsilon-\Lambda a_{0}^2)+b\right] \,, \label{a1} \\
	& a_{2} = +\frac{a_{0}}{c_{0}^2}\left[ (\epsilon-\Lambda a_{0}^2)^2
	+b \left(2 \epsilon + a_0^2 \left(\frac{1}{8 k} - \frac{7 \Lambda}{3} \right) \right) + b^2\right] \,,  \label{a2}\\
	&  c_1 = 2\epsilon - \Lambda a_0^2  + 3b \,,  \label{c1} \\
&  c_2 =	-  \frac{ 1}{6 k c_0 } \left[2 k \left(\epsilon ^2-\left(3 b - a_0^2 \Lambda +2 \epsilon
			\right)^2\right)+3 b a_0^2 \right] \,,   \\
& c_3 = \frac{a_0^4 b (3-8k\Lambda)}{96 k^2 c_0^2}\, .
	\end{align}

To summarize, the above solution contains three integration constants $a_0$, $c_0$, and $b$, along with the expansion radius $r_0=r_h$, the sign of the curvature of the transverse space $\epsilon$, and the constants of the theory $k$ and $\Lambda$. However, thanks to the gauge freedom \eqref{scalingfreedom}, the number of physical parameters boils down to {\em two} -- essentially, the mass and the Bach parameter.

We will show below in section~\ref{subsec_Einstein} that if $b=0$, then $B_{ab}=0$ everywhere (i.e., not just at the horizon), in which case the solution becomes Einstein (cf. also~section~\ref{subsec_QG}). The parameter $b$ thus measures how the solution departs from a (topological) Schwarzschild--(A)dS black hole and plays the role of a  gravitational (Bachian) ``hair'' (see also section~\ref{subsec_hairy} below).  Because of~\eqref{a_0}, \eqref{hor_conds} and \eqref{a1}, at an outer black-hole horizon, it will be constrained by
\be
	\epsilon-\Lambda a_{0}^2+b>0 \,.
	\label{hor_conds2}
\ee

From \eqref{Schwarz}, using \eqref{a1}--\eqref{c1}, the first two leading terms in the expansion of the metric functions $h(\bar r)$, $f(\bar r)$  around any horizon $\bar r_h=a_0$ read (assuming $a_1\neq0$, i.e., $ \epsilon-\Lambda a_{0}^2+b\neq0$)
\beqn
& & \hspace{-.4cm} h=\frac{a_0 c_0^2}{\epsilon-\Lambda a_{0}^2+b} \bar\Delta 
-\frac{c_0^2}{(\epsilon-\Lambda a_{0}^2+b)^3}
\left[ (\epsilon-\Lambda a_{0}^2)\epsilon
+b\left(3\epsilon-\frac{7}{3}\Lambda a_{0}^2+\frac{1}{8k}a_{0}^2
\right)+2b^2\right]\bar\Delta^2+\ldots \,,\label{h_Schw} \\
& & \hspace{-.4cm} f=\frac{\epsilon-\Lambda a_{0}^2+b}{a_0}\bar\Delta
-\frac{1}{a_{0}^2(\epsilon-\Lambda a_{0}^2+b)}
\left[(\epsilon-\Lambda a_{0}^2)\epsilon
+b\left(3\epsilon-\Lambda a_{0}^2-\frac{3}{8k} a_0^2
\right)+2b^2\right]\bar\Delta^2 +\ldots\,, \label{f_Schw}  
\eeqn
where $\bar\Delta=\bar r-\bar r_h$.
Note that in the gauge $a_1=a_0^2$ (then $c_0=-(\epsilon-\Lambda a_{0}^2+b)/a_0$ from \eqref{a1}), $f$ and $h$ coincide at the leading order, and $f=h$ in the limit $b\rightarrow 0$.

\subsection{Identifying the background Einstein spacetimes ($b=0$)}

\label{subsec_Einstein}

Let us now discuss the subclass of solutions for which the Bach parameter $b$ vanishes. As we will show below, this consists of two families of Einstein's spacetimes, for which the Bach tensor is necessarily zero. Because of~\eqref{hor_conds2}, solutions with a non-extremal black hole horizon must now satisfy
\be
	\epsilon-\Lambda a_{0}^2>0 \,.
	\label{hor_conds_Einst}
\ee
In particular, for $\Lambda>0$, only spherical black holes ($\epsilon=+1$) will be possible.

With $b=0$, the coefficients~\eqref{nonSchwinitcondc}, \eqref{nonSchwinitconda} reduce to 
\bea
&& a_i=a_0\,\bigg(\!-\frac{\epsilon -\Lambda a_0^2}{c_0}\bigg)^i
\qquad  \forall\ i \ge 0 \,,\quad
\\
&&  c_1=2\epsilon-\Lambda a_0^2 , \qquad c_2 = \frac{1}{3c_0} (\epsilon-\Lambda a_0^2) (3\epsilon-\Lambda a_0^2) \label{c2_back}
\,, 
\qquad c_i=0 \quad   \forall\ i \ge 3\,. 
\eea

Note, in particular, that the $c_i$ sequence is truncated as 
\be
	\H (r)= c_0(r-r_h)+c_1(r-r_h)^2+c_2(r-r_h)^3 \,, \label{IIbH0}
\ee
with~\eqref{c2_back}. There appear two possibilities depending on whether $\epsilon -\Lambda a_0^2=0$ or not.

\subsubsection{Generic case	$\epsilon -\Lambda a_0^2\not=0$: (A)dS-Schwarzschild metric}
	
\label{subsubsec_Einst_generic}

Assuming $\epsilon -\Lambda a_0^2\neq0$, the $a_i$ sequence is a geometric series, giving rise to
\be
	\Omega(r)= a_0 \,\sum_{i=0}^\infty  \,\Big(-(\epsilon-\Lambda a_0^2)\frac{\Delta}{c_0}\Big)^i
	=\frac{a_0\,c_0}{c_0+(\epsilon-\Lambda a_0^2)\Delta}=\frac{a_0\,c_0}{(\epsilon-\Lambda a_0^2)(r-r_h)+c_0}\,. \label{IIbOmega} 
\ee
Employing the gauge freedom \eqref{scalingfreedom}, we can set
\begin{equation}
	a_0 =-\frac{1}{r_h} \,, \qquad c_0 =r_h \epsilon - \frac{\Lambda}{r_h}  \label{IIb_a0}
\end{equation}
(which also means $a_1=a_0^2$), reducing the metric functions  to
\be
{\bar r}=\Omega(r) = -\frac{1}{r}\,, \qquad
\H (r) = \frac{\Lambda}{3} - \epsilon r^2 - \left(\frac{\Lambda}{3} - \epsilon r_h^2 \right) \frac{r^3}{r_h^3} \,. \label{IIbH0Schw}
\ee

Using~\eqref{physmet}, \eqref{Schwarz}, it is easy to see that this solution corresponds to the well-known \cite{Stephanibook,GriPodbook} (A)dS-Schwarzschild solution with a transverse space of arbitrary curvature, for which the only integration constant is usually rewritten as $2m=\left(\frac{\Lambda}{3} - \epsilon r_h^2 \right) \frac{1}{r_h^3}$,
and  the metric functions in the physical coordinates~\eqref{physmet} take the form  
			\be
			h({\bar r})=f({\bar r})^{-1}  = \epsilon - \frac{2 m}{{\bar r}}- \frac{\Lambda}{3} {\bar r}^2.
			\ee 
In addition to the standard spherical Kottler metric (for $\epsilon=+1$, cf., e.g., \cite{GibHaw77,HawPag83}), it describes topological black holes in Einstein's gravity \cite{Lemos95,HuaLia95,AmiBenHolPel96,Mann97,Vanzo97,BriLouPel97}. Here, the condition~\eqref{hor_conds_Einst} implies, for example, the known fact that black holes with $\epsilon\le0$ require $\Lambda<0$. When equality holds (i.e., $\epsilon-\Lambda a_{0}^2=0$) one obtains the standard extremality condition for Einstein black holes \cite{GibHaw77,Vanzo97,BriLouPel97}.\footnote{
Note indeed that, although we have obtained the closed form~\eqref{IIbH0Schw} of the solution under the assumption $\epsilon -\Lambda a_0^2\neq0$, the metric defined by~\eqref{IIbH0Schw} admits also the possibility $\epsilon -\Lambda a_0^2=0$, giving rise to a spacetime with a degenerate horizon at $r=r_h$ when $\Lambda\neq0$, and to a flat spacetime when $\Lambda=0$. In the usual form of the solution expressed in terms of $m$ (cf. above), one can further set $\Lambda=0=\epsilon$ while keeping $m\neq0$, which corresponds to a naked singularity.\label{footn_extremal}} In the case $\epsilon-\Lambda a_{0}^2<0$, the metric is time-dependent in the vicinity of the horizon in the exterior region (cf.~\eqref{h_Schw}, \eqref{f_Schw} with $b=0$) so that the hypersurface $\bar r=\bar r_h$ cannot be an outer black hole horizon (nevertheless, the spacetime can contain an outer black-hole horizon located elsewhere when $\epsilon\Lambda>0$, cf. \cite{GibHaw77,Vanzo97,BriLouPel97} for details).

\subsubsection{Special case $\epsilon -\Lambda a_0^2=0$: direct product Einstein spacetimes}
\label{sec_specialLim}

When $\epsilon -\Lambda a_0^2=0$ one simply obtains $a_i=0$ for all $i>0$, $c_1=\epsilon$ and $c_i=0$ for all $i>1$, i.e.,
\be
 \Omega= a_0 , \qquad \H= c_0(r-r_h)+\epsilon(r-r_h)^2 ,
\ee
which describes an Einstein spacetime of the  Kundt class in the form of a direct product metric of the {(anti-)}Nariai type (namely dS$_2\times$S$^2$, flat space, or AdS$_2\times$H$^2$, depending on the value of $\epsilon$). The horizon at $r=r_h$ is a Killing horizon, but not a black hole one (cf.~\eqref{hor_conds_Einst}). With a coordinate transformation (cf. \cite{Stephanibook}), one can always set $c_0=0=r_h$. This class of metrics is related to the near-horizon geometry of the extremal limits of the black holes of section~\ref{subsubsec_Einst_generic} (cf., e.g., the review \cite{KunLuc13} and references therein).

One might wonder why, although the field equations \eqref{Eq1C}, \eqref{Eq2C} can be solved exactly in the Einstein limit, we have not recovered here all Einstein spacetimes of the form~\eqref{BHmetric} (or \eqref{physmet}) satisfying $\epsilon -\Lambda a_0^2=0$. For example, in the $\epsilon=0=\Lambda$  case, such a solution is the  AIII  metric of \cite{EK} (see \cite{Stephanibook,GriPodbook} for further references and a physical interpretation) for which $\Omega \propto 1/r$ and  $\H \propto r^3$, which represents a naked singularity located at $r \rightarrow \infty$ (cf. also footnote~\ref{footn_extremal}). Such a type of solutions does not appear here because, in this section, we have considered the Einstein limit only of the [0,1] class -- which, by construction, contains a Killing horizon --  and thus horizonless metrics belonging to other cases in Table~\ref{table_cases} may not appear.

\subsection{More general solutions: black holes with nonvanishing Bach tensor}

\subsubsection{Generic case $\epsilon -\Lambda a_0^2\neq0$}

\label{subsec_hairy}

As mentioned above, the general quadratic-gravity solution of section~\ref{subsec_general} is non-Einstein when $b\neq0$. In this section, we will study a subset of black holes obeying \eqref{hor_conds2} which admit the (A)dS-Schwarzschild metric \eqref{IIbH0Schw} as a $b\rightarrow 0$ limit (cf. section~\ref{subsubsec_Einst_generic}). See the end of this section for a discussion of the Einstein limit for the distinct cases $\epsilon>\Lambda a_0^2$ and $\epsilon<\Lambda a_0^2$.

In order to express the metric functions $\Omega(r)$ and $\H(r)$ as the (A)dS-Schwarzschild background plus a quadratic-gravity correction (cf.~\eqref{Omega_[0,1]}, \eqref{H_[0,1]}), we reparametrize the series coefficients $a_i$ and $c_i$ by introducing coefficients $\alpha_i, \gamma_i$ as in \cite{Pravdaetal17,Svarcetal18,Podolskyetal18,Podolskyetal20}. Using again the gauge~\eqref{IIb_a0}, one obtains 
\bea
a_i \eqdef\tilde a_i-\frac{b}{r_h}\,\frac{\alpha_i}{(-r_h\rho)^i}\,,\qquad\hbox{where}\quad
\tilde a_i\equiv a_i(b=0)=\frac{1}{(-r_h)^{1+i}}  \ \ \ i\geq 0 \,,\label{def_alphai}\\
c_1 \eqdef 2\epsilon -\frac{\Lambda}{r_h^2} + 3b\,\gamma_1 ,\quad 
c_2 \equiv \frac{3 \epsilon r_h^2 - \Lambda}{3 r_h^3} +3b \, \frac{r_h\gamma_2}{\epsilon r_h^2-\Lambda},\ 
\quad c_i 
\equiv 3b\, \frac{\gamma_i \,r_h^{i-1}}{(\epsilon r_h^2-\Lambda)^{i-1}}
\ \ \quad i \ge 3 ,
\label{def_gammai}
\eea
where 
\begin{align}
	\X &\equiv  \epsilon -\frac{\Lambda}{r_h^2}\,, \label{rho_[0,1]}
\end{align}
and
\begin{align}
	\alpha_0\equiv 0\,, \quad \alpha_1 = 1 \,, \quad \gamma_1=1\,,\quad
	\gamma_2 = \frac{1}{3}\Big[4\epsilon - \frac{1}{r_h^2} \Big(2\Lambda+\frac{1}{2k}\Big) +3b\Big] \,. \label{alphasgammainitial_[0,1]}
\end{align}
The remaining coefficients $\alpha_i$ and $\gamma_i$ will be specified shortly.

The functions ${\Omega}$ and ${\H}$ then read
\begin{align}
	\Omega(r) & = -\frac{1}{r}-\frac{b}{r_h}\sum_{i=1}^\infty\alpha_i\Big(\,\frac{r_h-r}{\X\,r_h}\Big)^i \,, \label{Omega_[0,1]}\\
	\H (r) & = (r-r_h)\bigg[\,\epsilon \frac{r^2}{r_h} - \frac{\Lambda}{3r_h^3} \left (r^2+rr_h+r_h^2\right)
	+3b\,\X\,r_h\sum_{i=1}^\infty\gamma_i\Big(\,\frac{r-r_h}{\X\,r_h}\Big)^i\,\bigg] \,, \label{H_[0,1]}
\end{align}
indeed explicitly expressing the Bachian part of the metric as a correction to the (A)dS-Schwarzschild background, as desired.

Using \eqref{nonSchwinitcondc} and \eqref{nonSchwinitconda}, the  coefficients $\alpha_l, \gamma_{l+1}$ for ${l \ge 2}$ are given by the recurrent relations
\begin{align}
	&\alpha_{l}= \, \frac{1}{l^2}\Bigg[-\frac{2\Lambda}{3r_h^2}\,\sum_{j=0}^{l-1}\sum_{i=0}^{j}\left[\alpha_{l-1-j}\X^j+\big(\X^{l-1-j}
	+b\,\alpha_{l-1-j}\big)\big(\alpha_i \X^{j-i}+\alpha_{j-i}(\X^i+b\,\alpha_{i})\big)\right]
	\nonumber\\
	& \hspace{14.0mm}
	-\frac{1}{3}\alpha_{l-2}(2\epsilon+\X)\X(l-1)^2
	+\alpha_{l-1}\left[\frac{\epsilon}{3}+({\epsilon}+\X)\big(l(l-1)+\frac{1}{3}\big)\right]
	\nonumber\\
	& \hspace{14.0mm}
	-3\sum_{i=1}^{l}(-1)^i\,\gamma_i\,(\X^{l-i}+b\,\alpha_{l-i})\left[l(l-i)+\frac{1}{6}i(i+1)\right]\Bigg]\,, \nonumber\\
	&\gamma_{l+1}= \, \frac{(-1)^{l}}{kr_h^2\,(l+2)(l+1)l(l-1)}\sum_{i=0}^{l-1}\big[\alpha_i \X^{l-i}
	+\alpha_{l-i}\big(\X^i+b\,\alpha_i\big) \big](l-i)(l-1-3i) \, \quad  \forall\quad l\geq2\,.
	\label{alphasIIbgeneral}
\end{align}

Explicitly, the first few terms then read
\begin{align}
	& \alpha_2 =
	2\epsilon -\left(\frac{7}{3}\Lambda-\frac{1}{8k}\right)\frac{1}{r_h^2}+b
	\,, \nonumber\\
	& \alpha_3 =
	\frac{1}{9}\left[25\epsilon^2+\left(\frac{29}{8k}-\frac{179}{3}\Lambda\right)\frac{\epsilon}{r_h^2}+\left(\frac{1}{16k^2}-\frac{77}{24k}\Lambda+\frac{298}{9}\Lambda^2\right)\frac{1}{r_h^4}\right]\nonumber\\
	&\qquad+\frac{1}{9}\left[23\epsilon+\left(\frac{35}{8k}-\frac{104}{3}\Lambda\right)\frac{1}{r_h^2}\right]\,b+\frac{7}{9}\,b^2
	\,, \quad \ldots\,,
	\label{alphasIIb0}\\
	& \gamma_3 =
	\frac{1}{96k^2r_h^4}\left(1-\frac{8k}{3}\Lambda\right)
	\,, \nonumber\\
	& \gamma_4 =
	\frac{1}{18kr_h^2}\left[\frac{\epsilon^2}{5}+\left(-\frac{1}{4k}+\frac{4}{15}\Lambda\right)\frac{\epsilon}{r_h^2}-\frac{1}{160k^2r_h^4}-\frac{1}{45r_h^4}\left(14\Lambda^2-\frac{75}{8k}\Lambda\right)
	\right]\nonumber\\
	&\qquad+\frac{1}{720kr_h^2}\left[16\epsilon+\left(-\frac{13}{k}+\frac{56}{3}\Lambda\right)\frac{1}{r_h^2}\right]\,b
	+\frac{1}{90kr_h^2}\,b^2
	\,, \quad \ldots\,,
	\label{gammasIIb0}
\end{align}
which leads to
\begin{align}
	\Omega(r) & =-\frac{1}{r}+b\frac{(r-r_h)}{\X\,r_h^2}
	-\frac{b}{r_h}\left[ 2\X+\frac{1}{24 k r_h^2}(3-8k\Lambda)
	+b\right]
	\Big(\,\frac{r-r_h}{\X\,r_h}\Big)^2+\cdots
	\,,  \label{IIbOmegaFULL}\\
	\H (r) & = (r-r_h)\bigg\{\,\epsilon\frac{r^2}{r_h}-\frac{\Lambda}{3\,r_h^3}\left(r^2+rr_h+r_h^2\right)
	\nonumber \\
	&\hspace{18.0mm}
	+3b\,\X\,r_h\left[\Big(\,\frac{r-r_h}{\X\,r_h}\Big)+\frac{1}{3}\Big[4\epsilon-\frac{1}{r_h^2}\Big(2\Lambda+\frac{1}{2k}\Big)+3b\Big] \Big(\,\frac{r-r_h}{\X\,r_h}\Big)^2+\cdots\right]
	\,\bigg\}\,. \label{IIbH0FULL}
\end{align}

In the physical coordinates \eqref{physmet}, using \eqref{Schwarz} and the gauge
 \eqref{IIb_a0},
	 the first two leading terms in the expansion of the metric functions $h(\bar r)$, $f(\bar r)$  around the horizon $\bar r_h=a_0$ are (cf.~\eqref{h_Schw}, \eqref{f_Schw})
\beqn
f\!\!\!\!\!&=&\!\!\!\!\!\frac{\epsilon+b-\Lambda\bar r^2_h}{\bar r_h}\bar\Delta
-\frac{1}{\epsilon+b-\Lambda\bar r^2_h}
\left[ -\frac{3b}{8k}+(b+\epsilon)\frac{\epsilon+2b-\Lambda\bar r^2_h}{\bar r^2_h}\right]\bar\Delta^2
+\dots\,,\\
h\!\!\!\!\!&=&\!\!\!\!\!\frac{(\epsilon-\Lambda\bar r^2_h )^2}{{\bar r_h \left(\epsilon+b-\Lambda\bar r^2_h\right)}}\bar\Delta
-\frac{(\epsilon-\Lambda\bar r^2_h)^2}{\bar r^2_h\left(\epsilon+b-\Lambda\bar r^2_h\right)^3}
\left[ b\left(\frac{1}{8k}-\frac{7\Lambda}{3}\right) -\epsilon\Lambda+\frac{1}{\bar r^2_h}(\epsilon+b)(\epsilon+2b)\right]\bar\Delta^2+\dots\,.
\eeqn
Notice that $f\neq h$, except in the Einstein limit $b=0$. The first two orders of such an expansion were also displayed in \cite{Luetal12} using a different gauge.

Let us now discuss the Einstein limit in both cases $\epsilon -\Lambda a_0^2>0$ and $\epsilon -\Lambda a_0^2<0$ separately.
\begin{itemize}

	\item {\bf Case $\epsilon -\Lambda a_0^2>0$: (topological) Schwarzschild-Bach-(A)dS black holes}  \\
	In this case, in order to satisfy \eqref{hor_conds2},  $b$ is bounded from below
	by $\tilde b\equiv \Lambda a_0^2-\epsilon<0$. Thus we can approach $b\rightarrow 0$ from both sides while keeping the  inequality \eqref{hor_conds2} satisfied.
	Therefore, in this limit, the quadratic-gravity  black-hole horizon reduces to the (A)dS-Schwarzschild black-hole horizon and we will refer to them as {\em (topological) Schwarzschild-Bach-(A)dS black holes}. 
		
	\item {\bf Case $\epsilon -\Lambda a_0^2<0$: (topological) Schwarzschild-Bach-(A)dS black holes and  purely Bachian (topological) black holes} \\
	 In this case,  in order to satisfy \eqref{hor_conds2}, necessarily $b> \Lambda a_0^2-\epsilon>0$. Thus $b$ cannot approach $0$ without violating \eqref{hor_conds2} -- at some point, one reaches a critical value 	 $b_0= \Lambda a_0^2-\epsilon >0$ corresponding to $a_1=0$ and therefore
	 the metric functions $h(\bar r)$ and $f(\bar r)$ cannot be expressed
	 as power series in $\bar\Delta$ with integer powers, see \cite{Podolskyetal20,Pravdaetal21} for the case $\epsilon=+1$.
	 Nevertheless, in the Kundt coordinates, the expressions \eqref{Omega_[0,1]}, \eqref{H_[0,1]} still hold even for $b=b_0$ and the limiting procedure can be performed.   For all $0\leq b<b_0$, the horizon at $\bar r=\bar r_h$ is now a cosmological or an inner  horizon. The $b\rightarrow 0$ limit (cf. section~\ref{subsubsec_Einst_generic}) gives (A)dS-Schwarzschild  metric with a cosmological/inner horizon at $\bar r=\bar r_h$ which may or may not admit another (black-hole) horizon depending on the values of parameters $\epsilon$, $\Lambda$, and $r_h$. More precisely, the black-hole horizon can appear either for $\Lambda>0$ and $\epsilon>0$ or $\Lambda<0$ and $\epsilon<0$ with additional conditions on the parameters given in \cite{GibHaw77} and \cite{Vanzo97,BriLouPel97}, respectively. Einstein limits of these quadratic-gravity black holes are thus either (A)dS-Schwarzschild  black holes or naked singularities. If the limit is the (A)dS-Schwarzschild  black hole, we will refer to this black hole as a  {\em (topological) Schwarzschild-Bach-(A)dS black hole}. If the limit is a naked singularity, we will refer to this black hole a {\em purely Bachian (topological) black hole}. 

\end{itemize}

\begin{figure}[h!]
	\begin{center}
		\includegraphics[height=48mm]{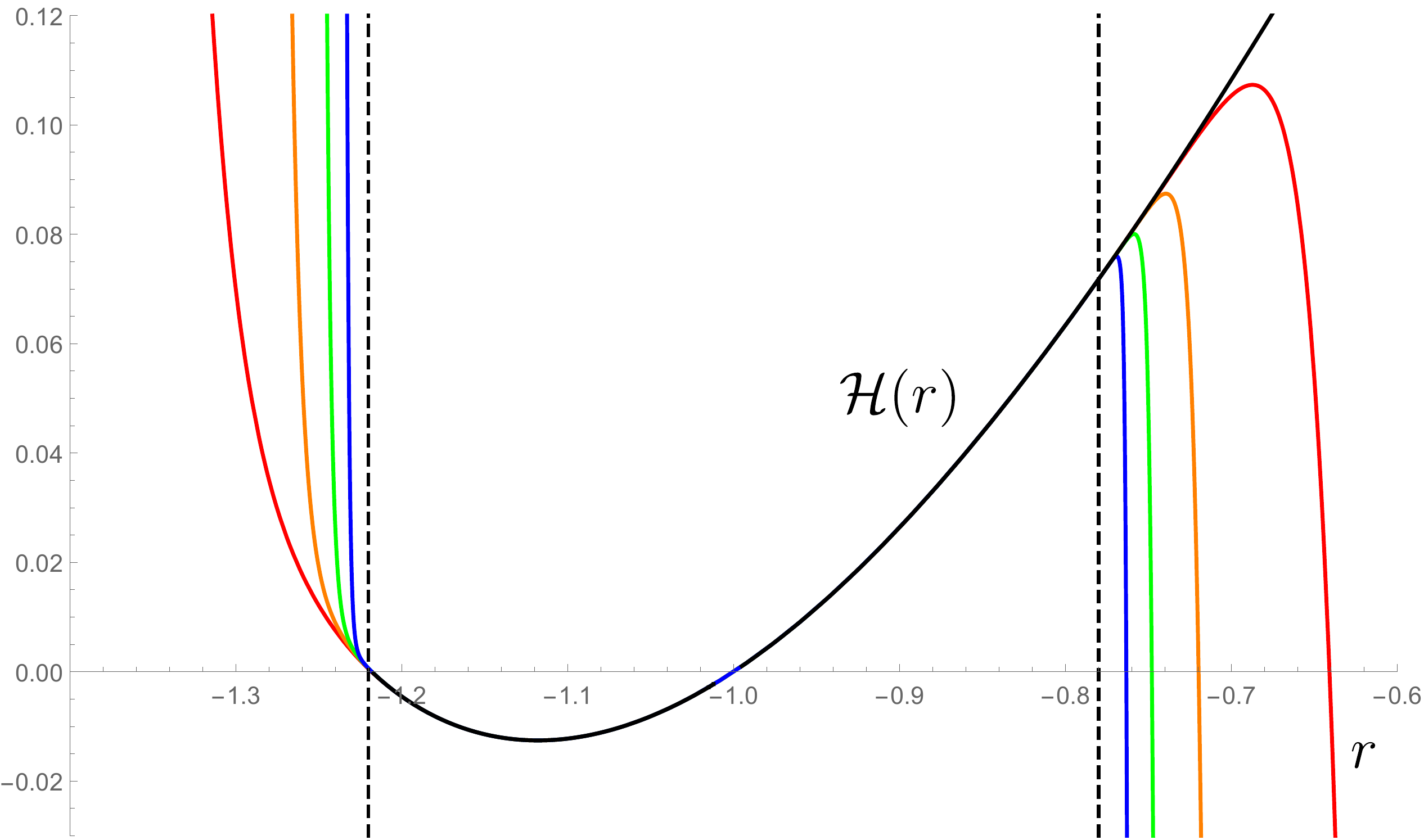}
		\hspace{5mm}
		\includegraphics[height=48mm]{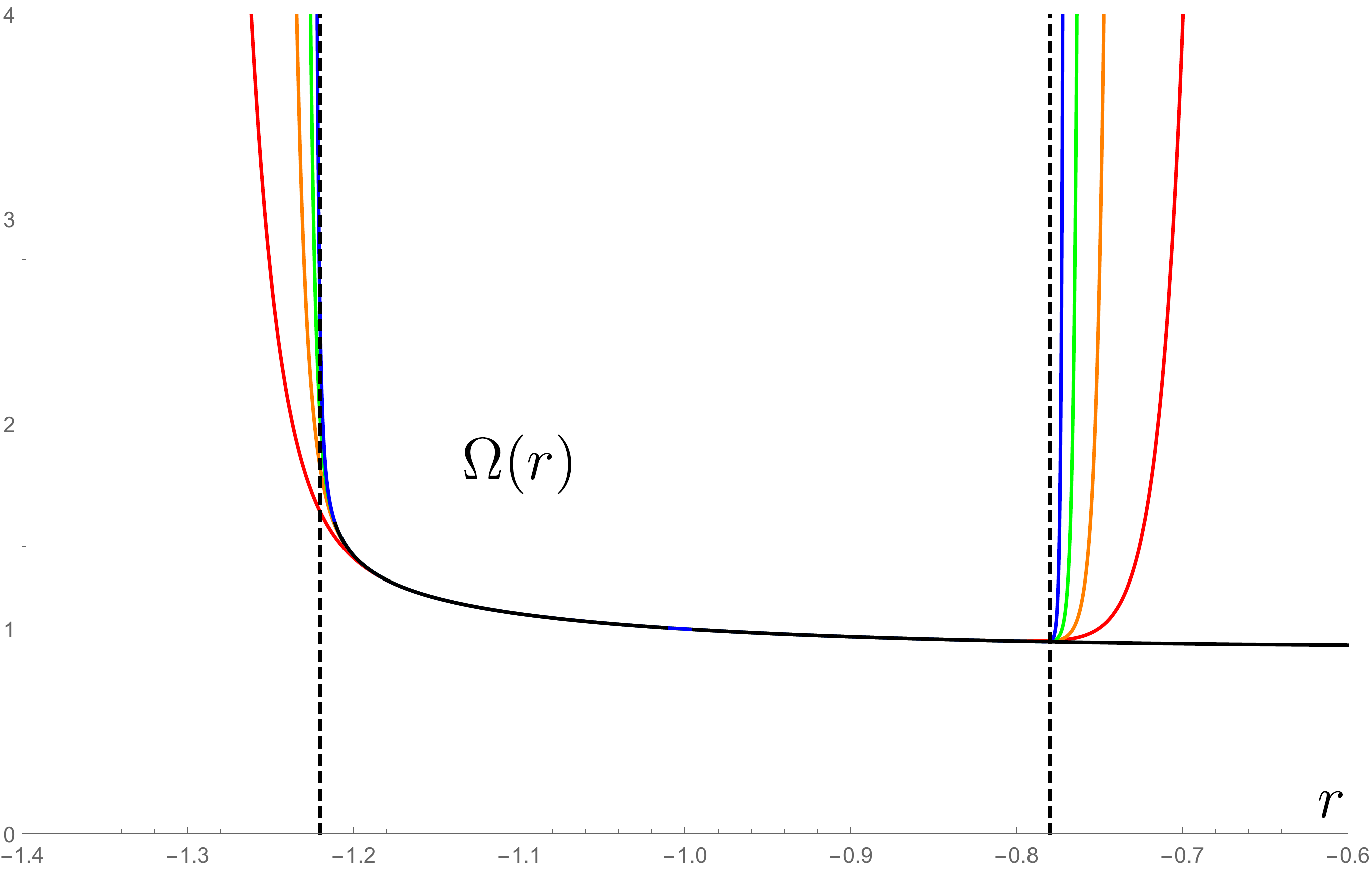}
		\caption{\label{fig:1a}   \small Series \eqref{Omega_[0,1]} and \eqref{H_[0,1]} for  the metric functions ${\cal H}(r)$ (left) and $\Omega(r)$ (right) for the first 20 (red), 50 (orange), 100 (green), and 200 (blue) terms and a numerical solution (black) of \eqref{Eq1C} and \eqref{Eq2C}  for values $\epsilon=0$, $\Lambda=0.2$, $r_h=-1$, $k=0.5$, $b=0.3$ (recall the gauge~\eqref{IIb_a0} is used here). From further analysis (not represented in the above figures, cf.~\cite{Svarcetal18,Pravdaetal21} in the case $\epsilon=+1$), the coefficients $\alpha_i$ and $\gamma_i$ seem to be approaching  geometric series for large values of $i$.	This allows us to estimate  the radius of the convergence and the interval of convergence indicated, in both graphs, by the two vertical dashed lines.  Note that, in the interval of convergence, $\bar r =\Omega(r)$ decreases as $r$ grows.  The horizon at $r=-1$ separates a static, outer region ($r<-1$, ${\cal H}<0$) from a time-dependent, inner one ($r>-1$, ${\cal H}>0$). While it is straightforward to estimate the lower bound of the interval of convergence in the physical coordinate ${\bar r}=\Omega(r)$ (using the right figure), it is difficult to estimate the upper bound, since it is determined by a (possible) intersection of  $\Omega(r)$ with the left vertical  dashed line. The numerical solution blows up precisely in the vicinity of the left vertical  dashed line, thus making an accurate estimate impossible from such a graph.   }
	\end{center}
\end{figure}

Let us conclude this section by presenting an example of a purely Bachian (toroidal) black hole with $\epsilon=0$, $\Lambda=0.2$, $r_h=-1$, $k=0.5$, $b=0.3$ in~Figure~\ref{fig:1a}, where the series approximation is depicted together with a numerical solution. Note that black holes of the form \eqref{physmet} with $\epsilon=0$ and $\Lambda>0$ are not allowed in Einstein's gravity.

\begin{figure}[h!]
	\begin{center}
		\includegraphics[height=60mm]{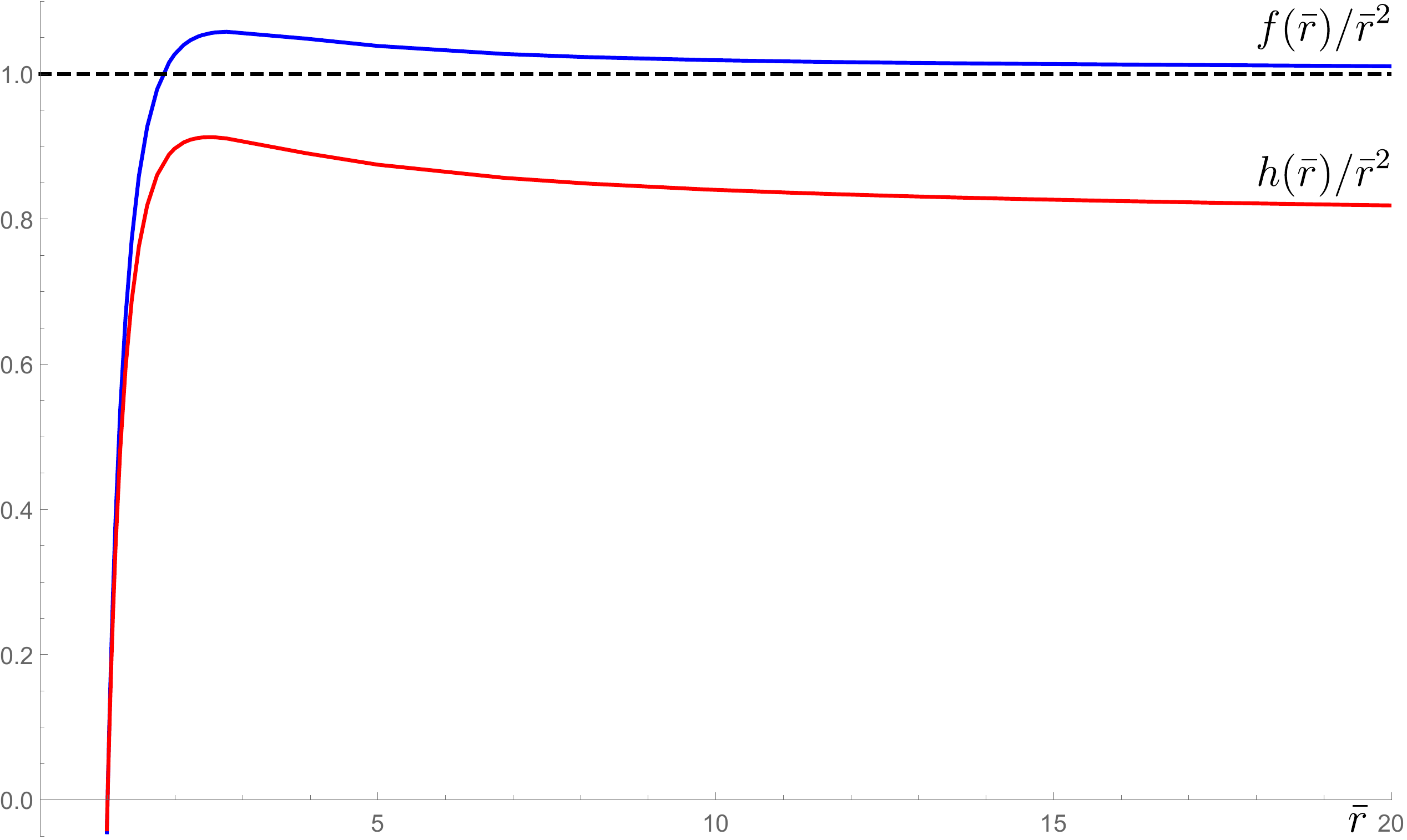}
		\caption{\label{fig:asympAdS}  Functions $f(\bar{r})/\bar{r}^2$ and $h(\bar{r})/\bar{r}^2$ for spherical Schwarzschild-Bach-AdS black holes obtained from the first 200 terms of series   \eqref{nonSchwinitcondc} and \eqref{nonSchwinitconda} for parameters  $r_h = -1$, $k = 1/3$, $b = 1/5$, $\Lambda = -3$, $\epsilon = 1$ (recall the gauge~\eqref{IIb_a0} is used here where, in general,  $h\not= f$ asymptotically; however, this could be remedied  by a gauge transformation $t\rightarrow \sigma t$). Further calculations suggest 	
			  that, also within a certain range of parameters of the solution, fine-tuning of the Bach parameter $b$ is not necessary to obtain an asymptotically  AdS  spacetime  (see section \ref{sec_concl} for further comments). This observation for the quadratic-gravity black holes  is in agreement with previous numerical results of \cite{Luetal12} obtained in the case of Einstein-Weyl gravity.   In contrast, in the $\Lambda=0$  case, fine-tuning is necessary (see Figure \ref{fig:finetune} and the numerical results of \cite{Luetal15,Luetal15b,Luetal17,Goldstein18,BonSil19,Huang22}). 
				}
	\end{center}
\end{figure}

Figure \ref{fig:asympAdS} shows that, within a certain continuous range of parameters, black holes of this section with $\Lambda<0$ are asymptotically AdS (here depicted for the spherical case $\epsilon=1$). This is in agreement with previous numerical results of \cite{Luetal12} obtained in the case of Einstein-Weyl gravity (see section \ref{sec_concl} for further comments).

\subsubsection{Special case $\epsilon -\Lambda a_0^2=0$}

\label{subsec_bachian}

In the special case $\epsilon -\Lambda a_0^2=0$, solutions $[0,1]$  with series expansions \eqref{rozvojcalH0} and coefficients $a_i$ and $c_i$ given in section~\ref{subsec_general} describe purely Bachian black holes provided condition \eqref{hor_conds2} is satisfied, i.e., $b>0$. This will be assumed in what follows.
	In the physical coordinates~\eqref{physmet}, 
 the first few orders of the metric functions  $h$ and  $f$ 
are given by \eqref{h_Schw} and \eqref{f_Schw}, respectively.

\begin{figure}[h!]
	\begin{center}
	\includegraphics[height=48mm]{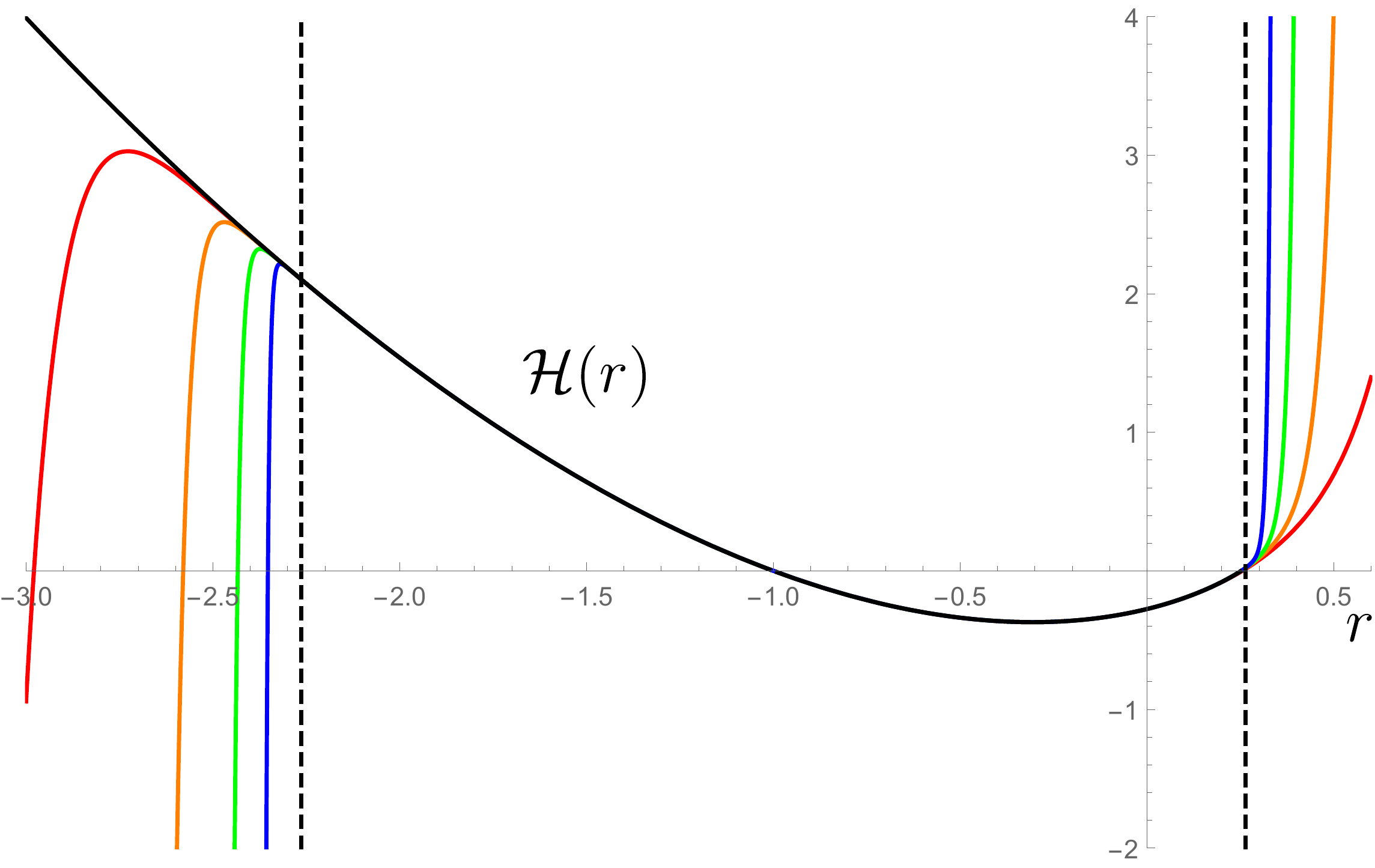}
	\hspace{8mm}
	\includegraphics[height=48mm]{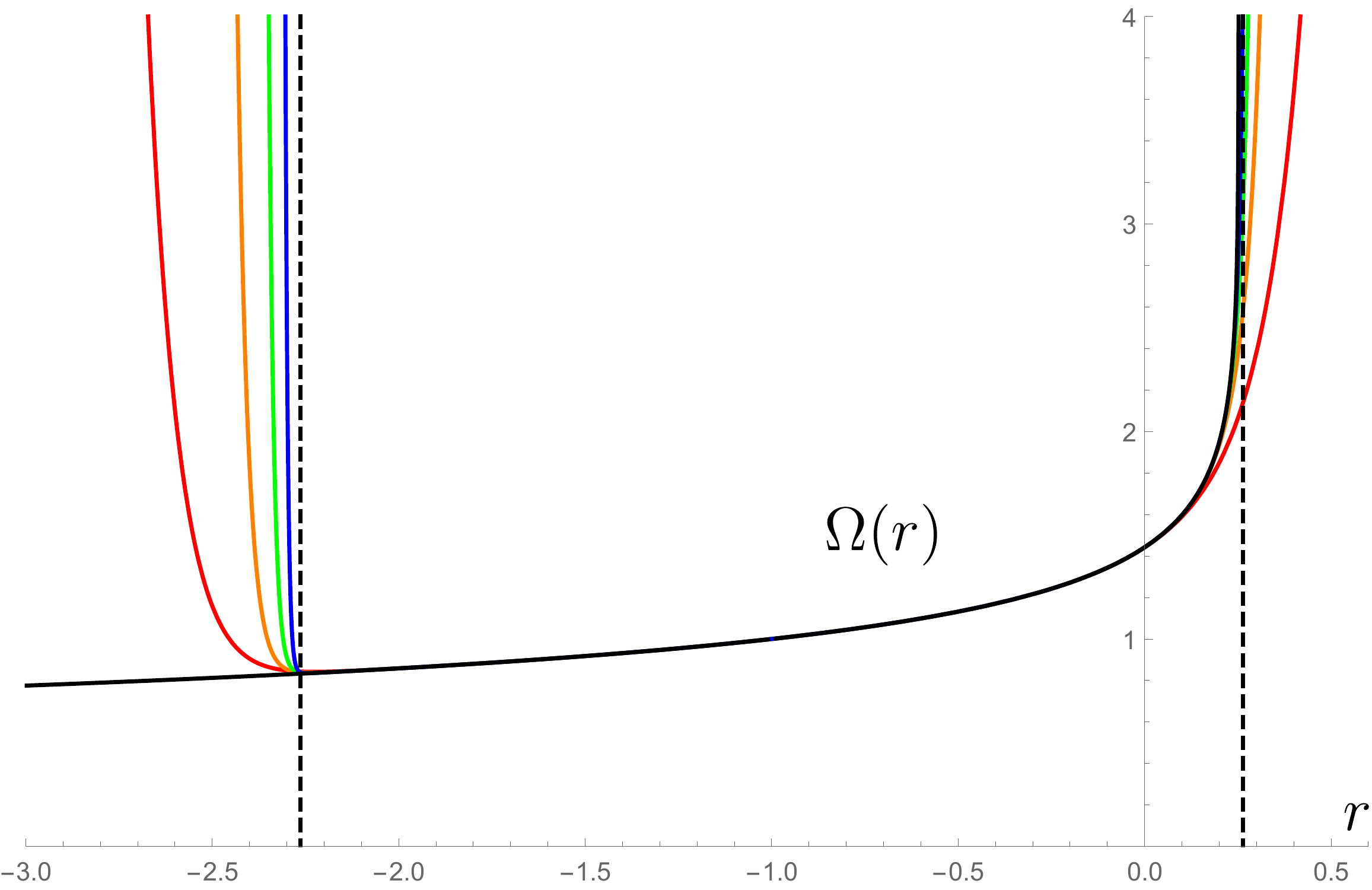}
	\caption{\label{fig:1}  \small Series \eqref{nonSchwinitcondc} and \eqref{nonSchwinitconda} for  the metric functions ${\cal H}(r)$ (left) and $\Omega(r)$ (right) for the first 20 (red), 50 (orange), 100 (green), and 200 (blue) terms for values $\epsilon=0$, $\Lambda=0$, $r_h=-1$, $k=1/2$, $c_0 = -1$, $a_0=1$, $b=1/5$. Similarly as in Fig.~\ref{fig:1a},
	the coefficients $a_i$ and $c_i$ seem to be approaching  geometric series for large values of $i$, which can be used to estimate the interval of convergence, denoted by the vertical dashed lines. The horizon at $r=-1$ separates a static, outer region ($r>-1$, ${\cal H}<0$) from a time-dependent, inner one ($r<-1$, ${\cal H}>0$).  
}
\end{center}
\end{figure}

In particular, within this special class, one can have black holes with $\epsilon=0=\Lambda$. By contrast, recall that Einstein's gravity with $\Lambda=0$ does not allow for
flat ($\epsilon=0$) black-hole horizons in vacuum \eqref{physmet} (cf.~\eqref{hor_conds_Einst}). 
 This is thus a new feature of quadratic gravity and planar horizons (and compactifications thereof) with $\Lambda=0$ are now allowed. For the special case of Weyl conformal gravity (i.e., $\gamma=0=\beta$), this was noted already in~\cite{Klemm98_cqg}. Coefficients $a_i$ and $c_i$ for these black holes can be obtained by substituting $\epsilon=0=\Lambda$ in \eqref{nonSchwinitcondc} and \eqref{nonSchwinitconda}. An example of such a solution is given in Figure~\ref{fig:1}.

\begin{figure}[h!]
      \begin{center}
           \includegraphics[height=48mm]{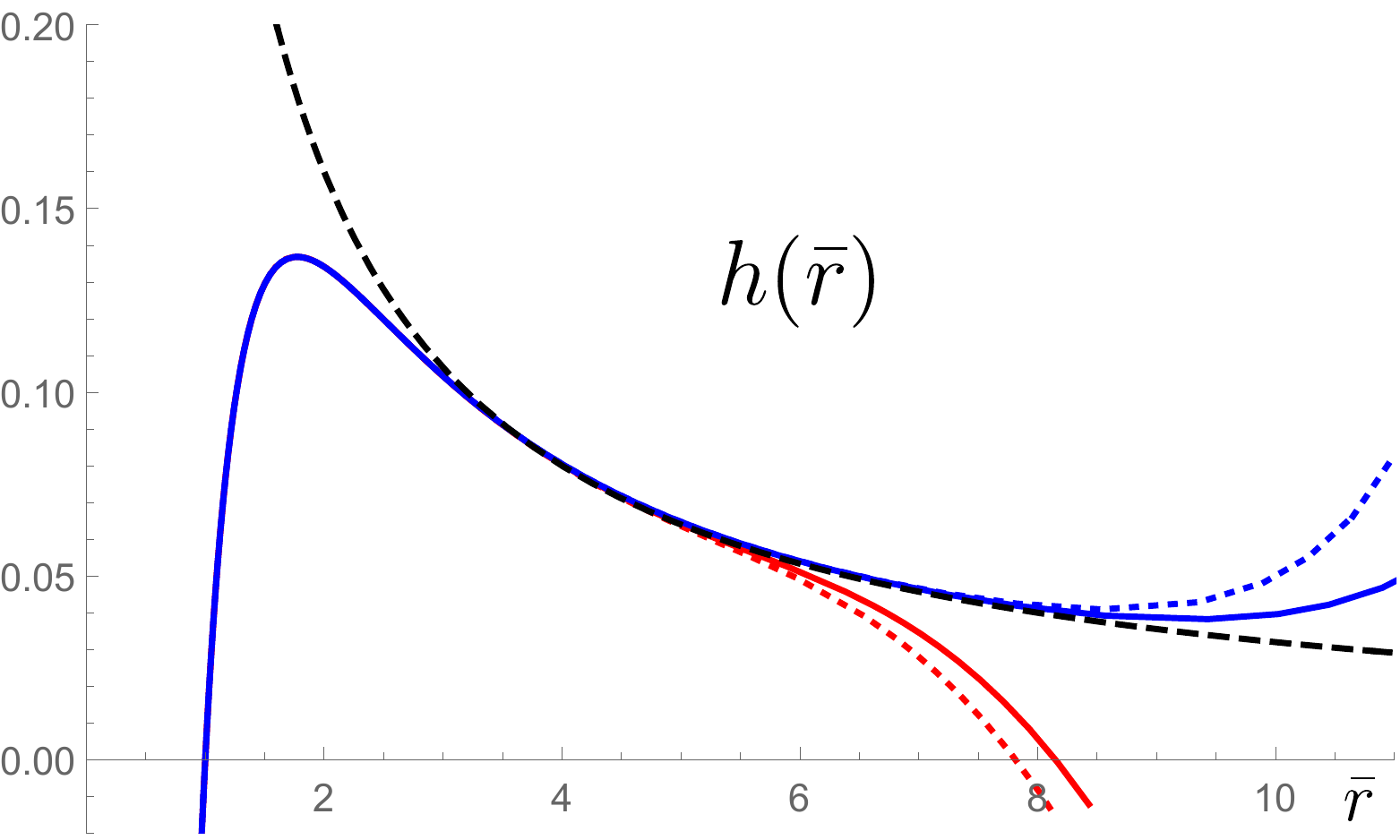}
				\caption{\label{fig:finetune}Function $h(\bar r)$ for   $b=1.575812$ (a blue dotted curve), $1.575811$ (blue), $1.5757$ (red), and  $1.5756$ (red dotted) and the general relativistic solution  $h(\bar r)=2\tilde m/{\bar r}$ for  $2\tilde m$=0.32 (black dashed) for values $\epsilon=0$, $\Lambda=0$, $r_h=-1$, $k=1/2$, $c_0 = -1$, $a_0=1$ (note that, in this gauge, $h\not= f$ asymptotically). Fine-tuning is done using the first 100 terms in the power-series expansions of the metric functions.}
       \end{center}
\end{figure}

 However, similarly as in the case of spherical horizons with $\Lambda=0$ \cite{Luetal15,Luetal15b,Luetal17,Goldstein18,Podolskyetal18,BonSil19,Podolskyetal20,Huang22}, it
turns out that, for generic values of the parameters $a_0$ and $b$, the metric functions $h$ and $f$ diverge as $\bar r\to+\infty$. In order to remedy this one needs to fine-tune the parameter $b$ (for any given $a_0$), such that in the weak gravity regime, one recovers a solution of general relativity. We have obtained the evidence that this is indeed possible by performing  fine-tuning using the first 100 terms in the series expansion of the solution (instead of using a numerical solution, as was done in \cite{Luetal15,Luetal15b,Luetal17,Goldstein18,BonSil19,Huang22}) -- this is shown in Figure~\ref{fig:finetune}. The asymptotic Ricci-flat spacetime (i.e., an AIII~metric \cite{EK,Stephanibook,GriPodbook}) is characterized by $h(\bar r)\propto f(\bar r)= 2m/{\bar r}$ (the parameter $m$ is determined by the parametres of the quadratic-gravity solution, $a_0$ and $b$; the equality $h=f$ could be achieved by an appropriate gauge transformation $t\rightarrow \sigma t$). A more precise approach to fine-tuning would be to match the expansion in the vicinity of the horizon with an asymptotic expansion in the form of logarithmic-exponential transseries (cf.~\cite{Goldstein18}) in the physical coordinates~\eqref{physmet}.

The Einstein limit of these spacetimes belongs to the  Kundt class in the form of a direct product metric of the {(anti-)}Nariai type (namely dS$_2\times$S$^2$, flat space, or AdS$_2\times$H$^2$, depending on the value of $\epsilon$, see section \ref{sec_specialLim}.

\section{Conclusions}

\label{sec_concl}

We have studied static black-hole solutions of the most general four-dimensional quadratic-gravity theory~\eqref{action} with a non-zero Einstein term ($\gamma\neq0$), under the assumption $R=$const (motivated by \cite{Luetal15b}). We have presented a solution  representing black holes possessing a non-extremal compact horizon of arbitrary topology. The solution is given in terms of an infinite power-series expansion (based on a Frobenius-like approach) around the horizon. Several different branches of the solution have been identified, which admit different Einstein limits -- accordingly, they can thus be interpreted as either higher-derivative ``corrections'' to Einstein black holes, or as purely Bachian black holes (for which the horizon does not survive the Einstein limit).

Although the general solution contains two independent integration constants (i.e., the black hole radius and the Bach parameter), for the special case of toroidal black holes with $\Lambda=0$, we have given evidence that solutions of physical interest (i.e., matching asymptotically an Einstein solution) need to be fine-tuned, such that there is in fact only one free parameter. This resembles corresponding results obtained using numerical methods in the case of spherical, asymptotically flat black holes \cite{Luetal15,Luetal15b,Luetal17,Goldstein18,BonSil19,Huang22}. Further investigation in this direction will be of interest and will play an important role in the study of  thermodynamics of these topological black holes. On the other hand, for theories with $\Lambda<0$ and $k>-3/(4\Lambda)$, numerical results of \cite{Luetal12} indicate that (for a given radius) there exists an open interval of values of $b$ such that these black holes are asymptotically AdS, with no need of any fine-tuning -- while the behavior becomes asymptotically Lifshitz at the extremes of that interval. This behaviour is  preserved also in full quadratic gravity, provided the Ricci scalar is constant (cf. eqs.~\eqref{eq:feq}, \eqref{k}, and Figure~\ref{fig:asympAdS}). The first law of the thermodynamics for these black holes was studied in~\cite{FanLu15} and contrasted with the asymptotically flat case \cite{Luetal15,Luetal15b,Luetal17,Goldstein18,BonSil19,Huang22}.

Finally, it is worth emphasizing that the simplified form of the field equations and the summary of classes of solutions provided in section~\ref{sec_FEQ} is of interest also in a broader context. Based on this, other types of solutions (such as extremal black holes, naked singularities, and wormholes) exist and will be studied elsewhere.

\section*{Acknowledgments}

This work has been supported by  the Czech Academy of Sciences (RVO 67985840)
 and research grant GA\v CR 19-09659S.

\bibliographystyle{unsrt}


\end{document}